\documentclass[sigconf,anonymous,review]{article}
\usepackage{spconf}

\usepackage{multirow}
\usepackage{multicol}
\usepackage{xcolor}
\usepackage{stfloats}
\usepackage{graphicx}
\usepackage{amsmath}

\usepackage{amssymb}
\usepackage{booktabs}
\usepackage{appendix}
\usepackage{float}
\usepackage[labelformat=simple]{subcaption}

\usepackage{mathtools}
\usepackage{url}
\usepackage{tabularx}
\usepackage{footmisc}
\usepackage{array}
\usepackage[flushleft]{threeparttable}
\usepackage{hyperref}
\usepackage{cleveref}

\usepackage{balance}

\crefname{section}{Sec.}{Secs.}
\crefname{table}{Tab.}{Tabs.}
\crefname{figure}{Fig.}{Figs.}

\newcommand{\eg}{\emph{e.g.,}~}

\newcommand{\wrt}{\emph{w.r.t.}~}

    \title{CLIPRerank: An Extremely Simple Method for Improving Ad-hoc Video Search}

\name{Aozhu Chen \qquad Fangming Zhou \qquad Ziyuan Wang \qquad Xirong Li\sthanks{Corresponding author: Xirong Li (xirong@ruc.edu.cn)}} 
 
\address{Renmin University of China\\ 
\href{https://github.com/ruc-aimc-lab/CLIPRerank}{https://github.com/ruc-aimc-lab/CLIPRerank}
}
\begin{document}
%
\maketitle

\begin{abstract}

Ad-hoc Video Search (AVS) enables users to search for unlabeled video content using on-the-fly textual queries. Current deep learning-based models for AVS are trained to optimize holistic similarity between short videos and their associated descriptions. However, due to the diversity of ad-hoc queries, even for a short video, its truly relevant part w.r.t. a given query can be of shorter duration. In such a scenario, the holistic similarity becomes suboptimal. To remedy the issue, we propose in this paper \textit{CLIPRerank}, a fine-grained re-scoring method. We compute cross-modal similarities between query and video frames using a pre-trained CLIP model, with multi-frame scores aggregated by max pooling. The fine-grained score is weightedly added to the initial score for search result reranking. As such, CLIPRerank is agnostic to the underlying video retrieval models and extremely simple, making it a handy plug-in for boosting AVS.  Experiments on the challenging TRECVID AVS benchmarks (from 2016 to 2021) justify the effectiveness of the proposed strategy. CLIPRerank consistently improves the TRECVID top performers and multiple existing models including SEA, W2VV++, Dual Encoding, Dual Task, LAFF, CLIP2Video, TS2-Net and X-CLIP.  Our method also works when substituting BLIP-2 for CLIP. 
\end{abstract}

\begin{keywords}
Ad-hoc video search, Large vision-language models, Video search reranking
\end{keywords}

\section{Introduction} \label{sec:intro}
\begin{figure} [thb!]
\centering
    \subfloat[Applied on winning solutions \label{fig:radar_winning}]{\includegraphics[width=0.5\columnwidth]{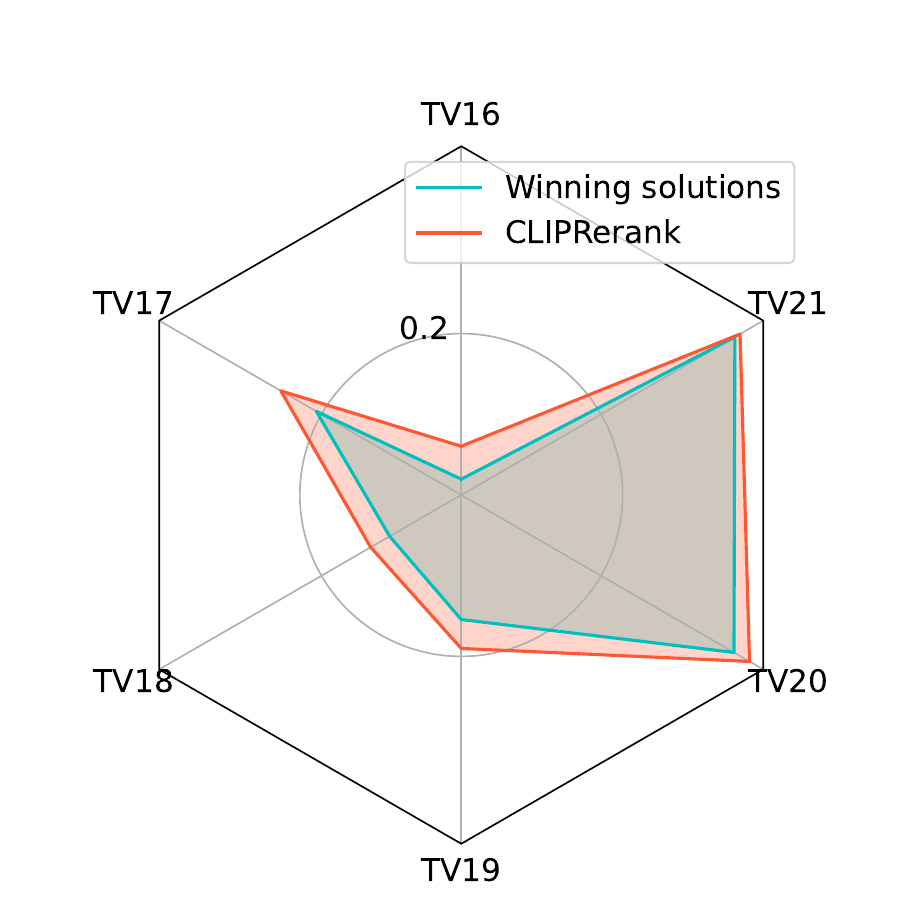}} 
    \subfloat[Applied on LAFF \label{fig:radar_laff}]{\includegraphics[width=0.5\columnwidth]{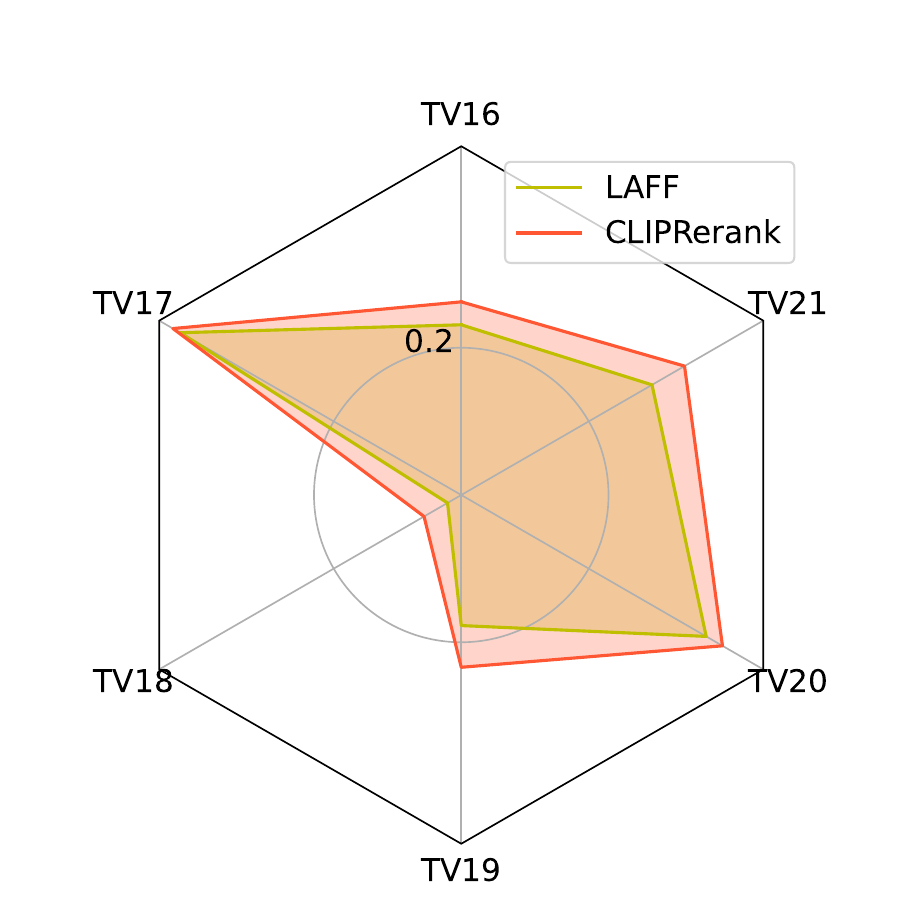}} 
    \caption{\textbf{Assessing CLIPRerank in the TRECVID AVS task}.}
    \label{Figure:radar}
\end{figure}
Ad-hoc video search (AVS) is fundamentally focused on creating a video search engine designed to enable everyday users to explore unlabeled short videos using natural language text queries. As a medium for information dissemination, the short video industry has experienced substantial growth in recent years. Concurrently, AVS has emerged as a compelling field situated at the nexus of natural language processing and computer vision. Existing text-to-video retrieval models can be categorized into two categories. The first category of these models uses multiple off-the-shelf text/visual features to re-learning a common space \cite{LiXirong2019W2VVPP, LiXirong2020SEA,wu2020interpretable_dual_task, Galanopoulos2020, Long_Mettes_Shen_Snoek_2020, cvpr19-zevr,hu2021lightweight_LAFF,xiang2023trust} to align text and video. Built upon the success of the Transformer architecture in natural language processing and Vision Transformer (ViT) as its generalization in computer vision \cite{vit2021}, another category of models emerges \cite{fang2021clip2video, xclip, mm22-negation-learning, liu2022ts2net}. These models leverage ViT-based visual encoders and Transformer-based text encoders to construct end-to-end solutions for text-to-video retrieval. In particular, the large pre-trained visual language model CLIP \cite{2021clip_icml} has demonstrated outstanding zero-shot performance across various downstream benchmarks, especially for enhancing the efficiency and accuracy of multimodal understanding. 

Since 2016, the annual TRECVID (TV) \cite{awad2016trecvid} evaluation has served as a pivotal benchmark for gauging advancements in the AVS task. Participants in this evaluation are tasked with developing video retrieval system's capable of retrieving the top 1,000 items for each test query from a vast collection of unlabeled short videos. The solutions for AVS most focus on inventing cross-modal video-text matching networks to align text and whole video by holistic similarity.
However, the limitation of holistic similarity becomes evident when dealing with the diverse nature of ad-hoc queries. In many cases, even within a short video, the segment that is truly relevant to a specific query may be considerably shorter. The holistic similarity metric, which considers the entire video in isolation, may lead to suboptimal results in such scenarios.

We propose in this paper CLIPRerank, an extremely simple method for improving AVS. In particular, the initial search results returned by a given video retrieval model are re-scored and consequently re-ranked based on CLIP-based frame-query similarities.  Though video search reranking is not new \cite{rerank_survey},  we see no attempt to apply reranking methods in the AVS task. As shown in \cref{Figure:radar}, the effectiveness of  CLIPRerank is assessed in the TRECVID AVS benchmark series,  effectively improving not only winning solutions of TV2016 to TV2021  but also state-of-the-art models.

\section{CLIPRerank: Re-scoring by CLIP} \label{sec:method}
Re-scoring plays a pivotal role in the context of text-to-video retrieval, primarily driven by the quest for enhanced retrieval performance. The initial retrieval results, although based on well-established models, may not fully capture the intricate nuances of semantic similarity between textual queries and visual content.
CLIP is a neural network trained on diverse web image-text pairs.
Due to the pre-training of a large-scale image-text corpus, it has a powerful visual-text modeling ability, which solves this 'training data-oriented' problem to a certain extent. 
Capitalizing on the strengths of both approaches, we introduce an extremely simple re-scoring method that contemplates employing CLIP to re-score cross-modal similarity to improve the original performance.

Suppose we have access to a top-ranked list of $k$ videos returned by a given video retrieval model $M$ \wrt a specific query $q$. For each video $v$ in the list, let $M(q, v)$ be the model-computed similarity score between $v$ and $q$. 
To re-rank the initial search results, we use CLIP to calculate the similarity between the query and the video's $i$-th frame $f_i$ as
\begin{equation}
    S(q,f_i) = \mbox{cosine}(TE(q), IE(f_i)),
\end{equation}
where $TE$ and $ IE$ indicate the text and image encoders of CLIP, respectively. The CLIP-based video-text similarity $S(q,v)$ is obtained by max pooling over the frame-level scores. 
Finally, through a weighted summation, the adjusted similarity score $S_{re}(q,v)$ is computed as
\begin{equation} \label{eq:true-sim}
S_{re}(q,v)= \alpha \cdot M(q, v) + (1 - \alpha) \cdot S(q,v),
\end{equation}
where $\alpha$ is a hyper-parameter that modulates the influence of each component. CLIPRerank technically differs from existing works that use CLIP directly for video-text matching \cite{ChenHu2021}, fuse CLIP features \cite{hu2021lightweight_LAFF} or re-train CLIP-based networks \cite{xclip}.

\section{Experiments} \label{sec:eval}

We investigate if CLIPRerank can improve the winning solutions of the TRECVID AVS task 2016-2021 (TV16-TV21) in the automated track. We also check if CLIPRerank works for current video retrieval models (that have not been evaluated on TRECVID). \\
\subsection{Experimental Setup}
\noindent \textbf{Test sets}. There are two test datasets: IACC.3 for TV16-TV18 and V3C1 for TV19-TV21, see \cref{tab:testdatasets}. 
IACC.3 contains approximately 4,600 Internet Archive videos with a mean duration of almost 7.8 minutes ~\cite{iacc2009}.
Through video segment boundary detection, these videos were divided into 335,944 short clips as the test set. 
V3C1 contains 7,475 videos from Vimeo with mean duration of almost 8 minutes \cite{V3C1Fabian2019}. Like IACC.3, these videos were divided into 1,082,659 short clips for testing. 
\begin{table}[tbh!]
\centering
\setlength{\abovecaptionskip}{0.cm}
\setlength{\belowcaptionskip}{-0.cm}
\normalsize
\caption{\textbf{Testsets used in TV16-TV21}. Frames are obtained by uniform sampling with a fixed time interval of 0.5 seconds.}
\label{tab:testdatasets}
\begin{center}
\scalebox{0.65}{
\begin{tabular}{lccccc}
\hline

\multirow{2}{*}{\textbf{Testset}}
&
\multirow{2}{*}{\textbf{Videos}}&\multirow{2}{*}{\textbf{Frames}}&\multirow{2}{*}{\textbf{Queries}}&\multicolumn{2}{c}{\textbf{Video length (s)}}   \\

\cline{5-6}
&&& &\textit{mean} & \textit{median}  \\
\hline
IACC.3 & 335,944 & 3,845,221 & 
\multicolumn{1}{c}{TV16: 30, TV17: 30, TV18: 30}   
& \multicolumn{1}{c}{7.8}  & \multicolumn{1}{c}{2.2}  \\ \hline
V3C1  & 1,082,649 & 7,839,450 &\multicolumn{1}{c}{TV19: 30, TV20: 30, TV21: 20}   &\multicolumn{1}{c}{3.3}  &\multicolumn{1}{c}{1.2} \\\hline
\end{tabular}

}
\end{center}
\end{table}
\begin{table*}[htbp!]
\caption{\textbf{Evaluating CLIPRerank on the TRECVID AVS benchmark series} (the automated track). }
\label{tab:compare_sota}
 \scalebox{0.78}{
\begin{tabular}{lllllllll}
\toprule
\multicolumn{1}{l}{\textbf{Model}} &\multicolumn{1}{l}{\textbf{CLIPRerank}}& \multicolumn{1}{l}{\textbf{TV16}} & \multicolumn{1}{l}{\textbf{TV17}} & \multicolumn{1}{l}{\textbf{TV18}} & \multicolumn{1}{l}{\textbf{TV19}}& \multicolumn{1}{l}{\textbf{TV20}} & \multicolumn{1}{l}{\textbf{TV21}} & \multicolumn{1}{l}{MEAN} \\ 
\hline
\multirow{2}{*}{Winning solutions}
& \multicolumn{1}{c}{-}
& \multicolumn{1}{l}{0.054 \cite{tv16-TOP1}} 
& \multicolumn{1}{l}{0.206 \cite{tv17-TOP1}} 
& \multicolumn{1}{l}{0.121  \cite{tv18-TOP1}} 
& \multicolumn{1}{l}{0.163  \cite{tv19-TOP1}} 
& \multicolumn{1}{l}{0.354  \cite{tv20-TOP1}} 
& \multicolumn{1}{l}{0.355  \cite{tv21-TOP1}}
& \multicolumn{1}{l}{---}  \\

& \multicolumn{1}{c}{+}
& 0.087 (61.1\%) 
& 0.247 (19.9\%) 
& 0.143 (18.2\%) 
& 0.192 (17.8\%) 
& 0.372 (5.1\%) 
& 0.361 (1.7\%) 
& ---  \\[2pt] \hline

\multirow{2}{*}{DualTask\cite{wu2020interpretable_dual_task}}
& \multicolumn{1}{c}{-}
& \multicolumn{1}{l}{0.185} 
& \multicolumn{1}{l}{0.241} 
& \multicolumn{1}{l}{0.123} 
& \multicolumn{1}{l}{0.185} 
& \multicolumn{1}{l}{---} 
& \multicolumn{1}{l}{---}
& \multicolumn{1}{l}{---} \\

& \multicolumn{1}{c}{+}
& 0.214  (15.7\%) 
& 0.277  (14.9\%) 
& 0.142  (15.4\%) 
& 0.210  (13.5\%) 
& --- 
& --- 
& --- \\[2pt] \hline

\multirow{2}{*}{W2VV++ \cite{LiXirong2019W2VVPP}}
& \multicolumn{1}{c}{-}
& \multicolumn{1}{l}{0.162} 
& \multicolumn{1}{l}{0.223} 
& \multicolumn{1}{l}{0.101} 
& \multicolumn{1}{l}{0.139} 
& \multicolumn{1}{l}{0.163} 
& \multicolumn{1}{l}{0.137} 
& \multicolumn{1}{l}{0.154}  \\
 
& \multicolumn{1}{c}{+}
& 0.204  (25.9\%)
& 0.260  (16.6\%) 
& 0.126  (24.8\%) 
& 0.168  (20.9\%) 
& 0.181  (11.0\%) 
& 0.160  (16.8\%) 
& 0.183  (18.8\%)
\\[2pt] \hline

\multirow{2}{*}{CLIP-zs \cite{ChenHu2021}}
& \multicolumn{1}{c}{-}
& \multicolumn{1}{l}{0.173}
& \multicolumn{1}{l}{0.202} 
& \multicolumn{1}{l}{0.092} 
& \multicolumn{1}{l}{0.124} 
& \multicolumn{1}{l}{0.134} 
& \multicolumn{1}{l}{0.197} 
& \multicolumn{1}{l}{0.154} \\

& \multicolumn{1}{c}{+}
& 0.170  (-1.7\%) 
& 0.209  (3.5\%) 
& 0.094  (2.2\%) 
& 0.124  (0\%)
& 0.136  (1.5\%) 
& 0.201  (2.0\%) 
& 0.155  (0.6\%) \\[2pt] \hline

\multirow{2}{*}{TS2-Net \cite{liu2022ts2net}}
&\multicolumn{1}{c}{-}
& \multicolumn{1}{l}{0.191}
& \multicolumn{1}{l}{0.245} 
& \multicolumn{1}{l}{0.112} 
& \multicolumn{1}{l}{0.120} 
& \multicolumn{1}{l}{0.153} 
& \multicolumn{1}{l}{0.188} 
& \multicolumn{1}{l}{0.168} \\

& \multicolumn{1}{c}{+}
& 0.196  (2.6\%)
& 0.258  (5.3\%) 
& 0.114  (1.8\%)
& 0.124  (3.3\%)
& 0.157  (2.6\%)
& 0.192  (2.1\%)
& 0.173  (3.0\%)\\[2pt] \hline

\multirow{2}{*}{DE \cite{cvpr19-zevr}} 
&\multicolumn{1}{c}{-}
& \multicolumn{1}{l}{0.163 } 
& \multicolumn{1}{l}{0.228 } 
& \multicolumn{1}{l}{0.116} 
& \multicolumn{1}{l}{0.164} 
& \multicolumn{1}{l}{0.186}
& \multicolumn{1}{l}{0.166} 
& \multicolumn{1}{l}{0.170 }  \\

& \multicolumn{1}{c}{+}
& 0.197  (20.9\%)	
& 0.267  (17.1\%)
& 0.133  (14.7\%)	
& 0.189  (15.2\%)	
& 0.207  (11.3\%)	
& 0.185  (11.4\%)	
& 0.196  (15.3\%) \\[2pt] \hline

\multirow{2}{*}{CLIP-FT \cite{hu2021lightweight_LAFF}} 
& \multicolumn{1}{c}{-}
& \multicolumn{1}{l}{0.191}
& \multicolumn{1}{l}{0.215} 
& \multicolumn{1}{l}{0.105} 
& \multicolumn{1}{l}{0.147} 
& \multicolumn{1}{l}{0.203} 
& \multicolumn{1}{l}{0.208} 
& \multicolumn{1}{l}{0.178}  \\

& \multicolumn{1}{c}{+}
& 0.189  (3.3\%)
& 0.236  (8.3\%) 
& 0.109  (1.9\%)
& 0.154  (7.7\%)
& 0.205  (2.0\%) 
& 0.213  (2.0\%)
& 0.184  (4.0\%)  \\[2pt] \hline

\multirow{2}{*}{X-CLIP \cite{xclip}} 
& \multicolumn{1}{c}{-}
& \multicolumn{1}{l}{0.209}
& \multicolumn{1}{l}{0.229} 
& \multicolumn{1}{l}{0.114} 
& \multicolumn{1}{l}{0.150} 
& \multicolumn{1}{l}{0.184} 
& \multicolumn{1}{l}{0.195} 
& \multicolumn{1}{l}{0.180}  \\

& \multicolumn{1}{c}{+}				
& 0.214  (2.4\%)
& 0.235  (2.6\%)
& 0.117  (2.6\%)
& 0.156  (4.0\%)
& 0.188  (2.2\%)
& 0.199  (2.1\%)
& 0.185  (2.8\%)  \\[2pt] \hline

\multirow{2}{*}{SEA \cite{LiXirong2020SEA} }
& \multicolumn{1}{c}{-}
& \multicolumn{1}{l}{0.153} 
& \multicolumn{1}{l}{0.235} 
& \multicolumn{1}{l}{0.129} 
& \multicolumn{1}{l}{0.169} 
& \multicolumn{1}{l}{0.201} 
& \multicolumn{1}{l}{0.199}  
& \multicolumn{1}{l}{0.181} \\

& \multicolumn{1}{c}{+}
& 0.196  (28.1\%)	
& 0.270  (14.9\%)	
& 0.149  (15.5\%)	
& 0.196  (16.0\%)	
& 0.223  (10.9\%)	
& 0.220  (10.6\%)	
& 0.209  (15.5\%) \\[2pt] \hline

\multirow{2}{*}{CLIP2Video \cite{fang2021clip2video} }
& \multicolumn{1}{c}{-}
& \multicolumn{1}{l}{0.176} 
& \multicolumn{1}{l}{0.229} 
& \multicolumn{1}{l}{0.114} 
& \multicolumn{1}{l}{0.176} 
& \multicolumn{1}{l}{0.207} 
& \multicolumn{1}{l}{0.255} 
& \multicolumn{1}{l}{0.193}  \\
& \multicolumn{1}{c}{+}
& 0.186  (5.7\%) 
& 0.242  (5.7\%)
& 0.119  (4.4\%) 
& 0.187  (6.3\%) 
& 0.214  (3.4\%) 
& 0.264  (3.5\%) 
& 0.202  (4.7\%)  \\[2pt] \hline

\multirow{2}{*}{LAFF \cite{hu2021lightweight_LAFF}} 
& \multicolumn{1}{c}{-}
& \multicolumn{1}{l}{0.211} 
& \multicolumn{1}{l}{0.285} 
& \multicolumn{1}{l}{0.137} 
& \multicolumn{1}{l}{0.192} 
& \multicolumn{1}{l}{0.265} 
& \multicolumn{1}{l}{0.235} 
& \multicolumn{1}{l}{0.221}  \\

& \multicolumn{1}{c}{+}
& 0.216  (2.5\%) 
& 0.293  (2.8\%) 
& 0.149  (8.9\%) 
& 0.194  (1.2\%) 
& 0.266  (0.3\%) 
& 0.236  (0.3\%)
& 0.226  (2.1\%)  

\\[2pt] \hline

\multirow{2}{*}{LAFF* }
& \multicolumn{1}{c}{-}
& \multicolumn{1}{l}{0.262} 
& \multicolumn{1}{l}{0.357} 
& \multicolumn{1}{l}{0.192} 
& \multicolumn{1}{l}{0.243} 
& \multicolumn{1}{l}{0.358} 
& \multicolumn{1}{l}{0.361} 
& \multicolumn{1}{l}{0.296}  \\

& \multicolumn{1}{c}{+}										
& 0.282  (7.6\%)
& 0.368  (3.1\%)
& 0.197  (2.6\%)
& 0.255  (4.9\%)
& 0.361  (0.8\%)
& 0.365  (1.1\%)
& 0.305  (3.1\%)  \\

\bottomrule
\end{tabular}
}
\end{table*}

\noindent\textbf{Video retrieval models}. Subject to the availability of a model’s PyTorch code, we collect eight models, five of which are based on off-the-shelf features (W2VV++, SEA, DualTask, DE and LAFF) and the other three are end-to-end trained (CLIP2Video, X-CLIP and TS2-Net). \\
$\bullet$ W2VV++ \cite{LiXirong2019W2VVPP}: It encodes a query with three parallel text encoders and the outputs are combined into one vector and mapped to a common space via an MLP. Similarly, the video feature is projected into this common space using an FC layer.\\
$\bullet$  SEA \cite{LiXirong2020SEA}: It leverages several text encoders within a multi-space framework, with each encoder aligned to a distinct common space and then averaging the similarities calculated within each space as video-text similarity .\\
$\bullet$  DE \cite{cvpr19-zevr}: Two multi-level encoding networks with similar architectures, one for queries and the other for videos. \\
$\bullet$  DualTask \cite{wu2020interpretable_dual_task}: It aims to improve the performance of video retrieval by associating embeddings with semantic concepts, making the search results more interpretable. \\
$\bullet$  LAFF \cite{hu2021lightweight_LAFF}: A lightweight attention-based feature fusion model, it conducts feature fusion by initially converting each of the $k$ features into a $d$-dimensional feature vector and subsequently aggregating these transformed features into a unified feature through a convex combination.
\\
$\bullet$  CLIP2Video \cite{fang2021clip2video}: It comprises two blocks, one for capturing detailed temporal dynamics in video frames, and the other for aligning video clip tokens with text phrases.\\
$\bullet$  X-CLIP \cite{xclip}: Computing multi-granularity similarities between text (sentence / words) and (video / frames). \\
$\bullet$  TS2-Net \cite{liu2022ts2net}: Dynamically alter visual token sequences and identify crucial tokens in temporal / spatial dimensions.

Additionally, we test the original CLIP (denoted as CLIP-zs \cite{ChenHu2021}) and a fine-tuned edition CLIP-FT \cite{hu2021lightweight_LAFF}.

\noindent \textbf{Evaluation criterion}.  
 We adopt the official metric, inferred Average Precision (infAP) \cite{2023trecvidawad}, and assess overall performance by averaging infAP scores over the given queries.

\noindent \textbf{Implementation details}.
For DE, W2VV++ and SEA, we follow the original papers, using ResNeXt-101\footnote{\url{https://github.com/xuchaoxi/video-cnn-feat}\label{cnn-feat}} and ResNet-152\footnote{\url{https://mxnet.apache.org/versions/1.0.0/tutorials/python/predict_image.html}} as visual features.
For X-CLIP and TS2Net, we sample 12 frames per video and use CLIP-B/32 as the visual backbone.
For a fair comparison, we train all models on MSR-VTT (9k training videos) \cite{msrvtt}. CLIP-B/32 is used for re-scoring, unless otherwise specified. 
Since we can only get the top 1k retrieved results per TRECVID run, to maintain consistency, the number of videos $k$ of the initial ranking list is also set to 1k. The weight $\alpha$ is $0.4$.
We run all experiments with PyTorch on two NVIDIA GeForce RTX 3090 GPUs.

\subsection{Results}

\textbf{The influence of CLIPRerank}. 
As  \cref{tab:compare_sota} shows, the inclusion of CLIPRerank improves the performance of all the models evaluated, with the most substantial enhancement reaching an impressive 61.1\% (from $0.054$ to $0.087$), as observed in the case of the TV16 winning solution.  In addition, for models like DualTask, W2VV++, DE, and SEA, which rely solely on pre-trained visual features, their performance improvements all exceeded 10\%. Even for LAFF, which already used CLIP as one of its feature extractors, we still achieve a relative improvement of 2.1\%. Similar results can also be observed on TS2-Net, CLIP-FT, X-CLIP, and CLIP2Video.
For example, CLIP2Video has shown an increase from 0.193 to 0.202 in overall performance on TV16-TV21, marking a performance improvement of 4.7\%. The experimental results allow us to conclude that  CLIPRerank improves the AVS performance.

\begin{figure}[tbp!]
    \centering
    \includegraphics[width=\columnwidth]{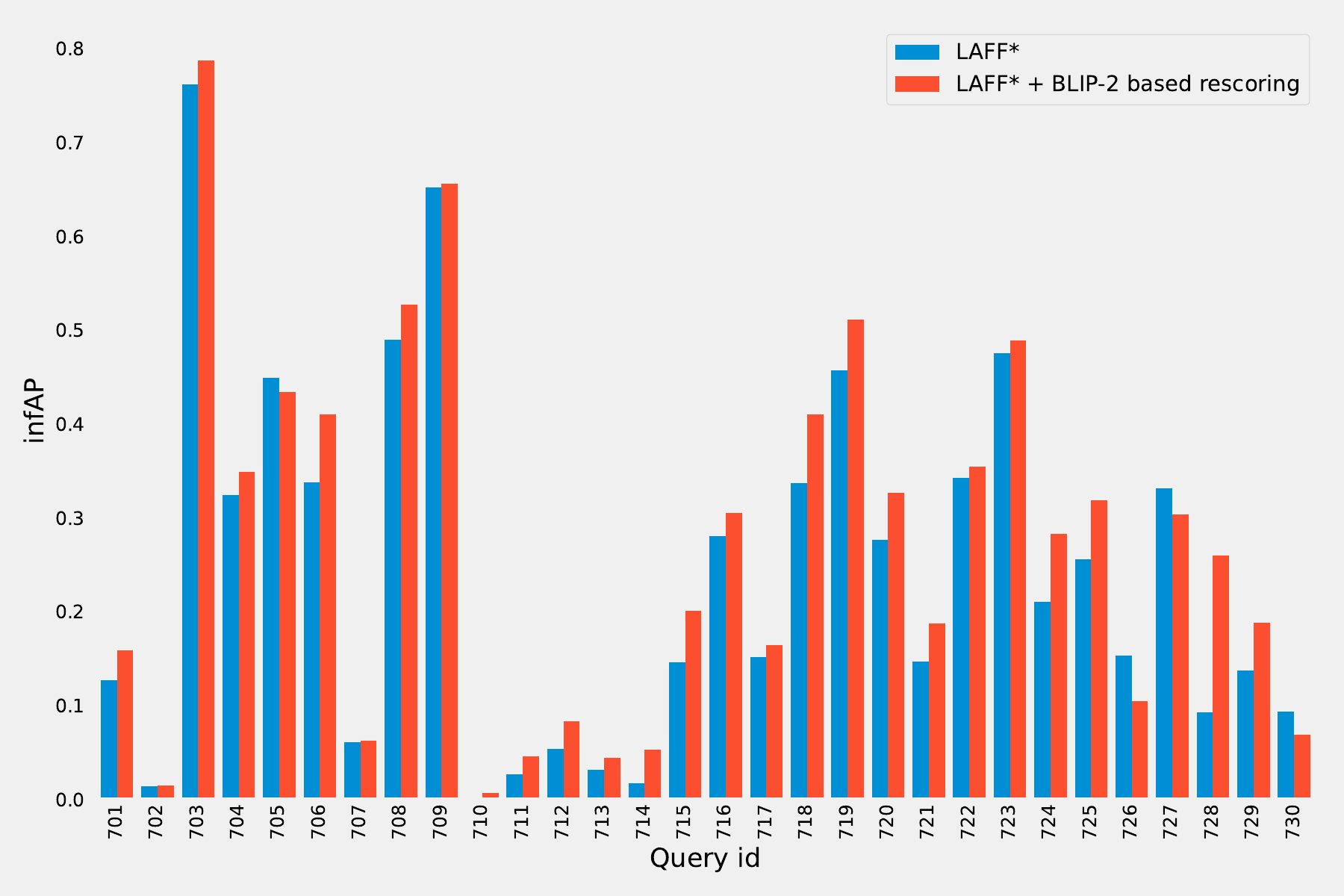}
    \caption{\textbf{ Per-query analysis on TV22}. We use the same experimental setups as LAFF$^*$ to test on the latest V3C2 test set with queries of TV22. BLIP-2 is used for re-scoring.}
    \label{fig:tv22}
\end{figure}
\begin{figure*}[bp]
    \centering
    \includegraphics[width=1.65\columnwidth]{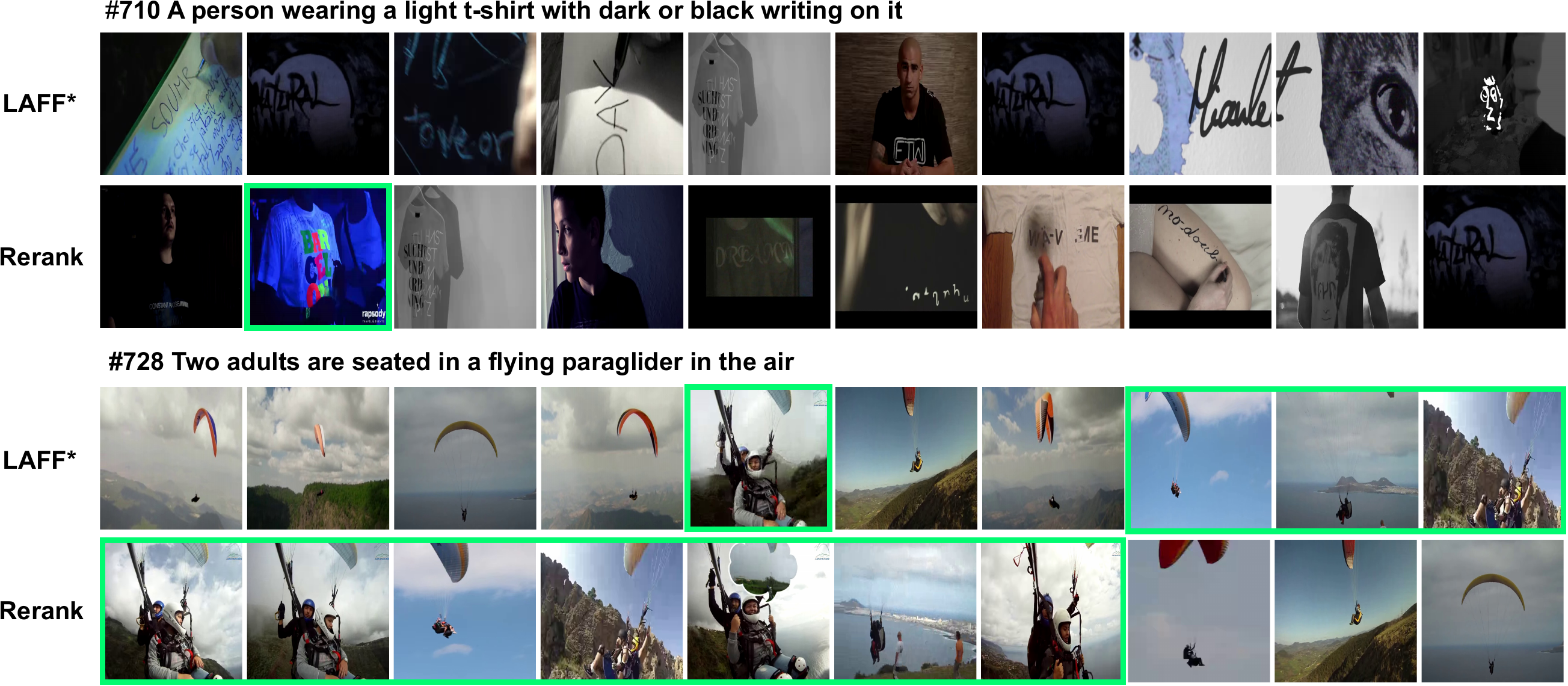}
    \caption{\textbf{Top-10 video search results by LAFF* and LAFF* + CLIPRerank, respectively}. Queries selected from TV22.}
    \label{fig:demo}
\end{figure*}

\textbf{CLIPRerank for stronger models}. To test  whether more powerful models can lead to better performance, we follow \cite{li2022renmin} to train a stronger version of LAFF, denoted as LAFF*. Moreover, 
we utilize BLIP-2\footnote{\url{https://github.com/salesforce/LAVIS/tree/main/projects/blip2}}, a more powerful Vision Language (VL) model based on CLIP, for reranking LAFF* on the V3C2 video dataset \cite{v3c2}, which is the test set of TV22. Specifically, given $k$ of 5k and  $\alpha$ of 0.5, the performance is increased from 0.241 to 0.271. Per-query analysis on TV22 shows that there remain difficult queries that the re-scoring fails to respond to, see \cref{fig:tv22}. It is noteworthy that a quite challenging query, \textit{\#710: A person wearing a light t-shirt with dark or black writing on it}, initially exhibited an infAP of 0.0002 for the initial result. However, after applying reranking, this metric increased to 0.007. Visualization of the retrieval results reveals that the original top 10 videos did not include any correct matches. Nevertheless, the effectiveness of the retrieval notably improved following the reranking, see \cref{fig:demo}, with a correct video being ranked second. On the other hand, we see that on query \textit{\#728}, the rerank performance improvement is very significant. It can be seen from \cref{fig:demo} that the video with "two adults" clearly appearing in the frame appeared in the front ranking. It indicates the excellent representation of the frame by the VL model can complement the retrieval model with significant static information.

\textbf{Comparison with existing reranking method}. As aforementioned, we see no attempt to apply reranking methods for AVS. Existing methods for video search reranking are mostly not open-source. We tried the classical LabelSpreading \cite{zhou2003learning}, performing semi-supervised label propagation on the initial results of LAFF* on TV16-TV21. Even with its all hyperparameters carefully tuned, LabelSpreading, with a mean performance of 0.294, does not excel LAFF*, see \cref{tab:compare_sota}. 

\textbf{Computational overhead}. We use CLIPRerank to rescore the top 5k results retrieved by a baseline model. Given features cached in memory, our Python implementation takes 22 ms per query. The overhead is insignificant.

\section{Conclusions} \label{sec:concs}
Our experiments on the challenging TRECVID AVS benchmarks, spanning from 2016 to 2021, demonstrate the efficacy of the proposed CLIPRerank method. Concerning the use of pre-trained large vision-language models (LVLM), \eg CLIP and BLIP-2, for text-to-video retrieval, our major finding is the following: Using an LVLM for search result reranking is better than using it directly for video-text matching. Our work highlights the potential for LVLM based fine-grained re-scoring, which matches significant static information in videos with relevant portions of text, thereby compensating for shortcomings in holistic similarity.  
The extreme simplicity of CLIPRerank and its model-agnostic nature make it a valuable and easy-to-use tool for improving ad-hoc text-to-video retrieval.





\medskip
\textbf{Acknowledgements}. The authors thank George Awad for sharing TRECVID AVS submissions and Jiaxin Wu for sharing the DualTask results. This research was supported by National Natural Science Foundation of China (No. 62172420) and Tencent Marketing Solution Rhino-Bird Focused Research Program.

\bibliographystyle{IEEEbib-abbrev}
\bibliography{strings,main}
\balance

\end{document}